\documentclass[aps,pra,superscriptaddress,floatfix,showpacs,10pt,onecolumn]{revtex4}
\usepackage{amsmath,amsfonts,amssymb,graphics,graphicx,epsfig,color,times,bbm,subfigure}

\begin{document}
\title{Generating multi-photon W-like states for perfect quantum teleportation and superdense coding}

\author{Ke Li}
\affiliation{School of Physics {\&} Material Science, Anhui University, Hefei 230601, China}

\author{Fan-Zhen Kong}
\affiliation{School of Physics {\&} Material Science, Anhui University, Hefei 230601, China}
\affiliation{Department of Computer Science, Jining University, Qufu, Shandong 273155, China}

\author{Ming Yang \footnote{mingyang@ahu.edu.cn}}
\affiliation{School of Physics {\&} Material Science, Anhui University, Hefei 230601, China}

\author{Fatih Ozaydin \footnote{mansursah@gmail.com}}
\affiliation{Department of Information Technologies, Isik University, Sile, Istanbul, 34980, Turkey}

\author{Qing Yang}
\affiliation{School of Physics {\&} Material Science, Anhui University, Hefei 230601, China}

\author{Zhuo-Liang Cao}
\affiliation{School of Electronic and Information Engineering, Hefei Normal University, Hefei 230601, China}
\affiliation{School of Physics {\&} Material Science, Anhui University, Hefei 230601, China}

\begin{abstract}
An interesting aspect of multipartite entanglement is that for perfect teleportation and superdense coding, not the maximally entangled W states but a special class of non-maximally entangled W-like states are required. Therefore, efficient preparation of such W-like states is of great importance in quantum communications, which has not been studied as much as the preparation of W states. In this letter, we propose a simple optical scheme for efficient preparation of large-scale polarization based entangled W-like states by fusing two W-like states or expanding a W-like state with an ancilla photon. Our scheme can also generate large-scale W states by fusing or expanding W or even W-like states. The cost analysis show that in generating large scale W states, the fusion mechanism achieves a higher efficiency with non-maximally entangled W-like states than maximally entangled W states. Our scheme can also start fusion or expansion with Bell states, and it is composed of a polarization dependent beam splitter, two polarizing beam splitters and photon detectors. Requiring no ancilla photons or controlled gates to operate, our scheme can be realized with the current photonics technology and we believe it enables advances in quantum teleportation and superdense coding in multipartite settings.
\end{abstract}



\maketitle

\section{introduction}
Quantum teleportation and superdense coding are two intriguing tasks in quantum information processing. In quantum teleportation, one can \lq \lq teleport \rq \rq an intact unknown state from one place to another by using a pre-shared bipartite entangled state, the so-called EPR pair and sending two bits of classical information~\cite{bennett1}.
A pre-shared EPR pair also enables superdense coding, which can double the classical capacity of a communication channel~\cite{bennett}.
However, due to inevitable environmental effects, a maximal EPR pair cannot be always available and it was shown that non-maximal EPR pairs cannot enable perfect quantum teleportation~\cite{akpati02,akpati04} and perfect superdense coding~\cite{hausladen, hao, pati}.
As the number of entangled particles increases, interesting states arise, which cannot be transformed into each other by stochastic local operations and classical communications (SLOCC) in general~\cite{bennett2} with the exceptions of three-partite states in the asymptotic regime~\cite{GHZ-W1,GHZ-W2}.
Multipartite entangled states can be used as resource for various quantum information and communication tasks and there are tasks that can be achieved with only a specific state \cite{hondt}.
W states~\cite{dur} form an important class of multipartite entanglement, with a robust structure against particle losses~\cite{dur01}.
Therefore, in contrast to GHZ states~\cite{GHZ}, W states can still be used as a resource even after the loss of particles.
Recently, the preparation schemes and applications of maximally entangled W states have attracted considerable attentions~\cite{tashima, ozdemir1, yesilyurt, bugu, ozdemir2, ozaydin, ozaydin1, yesilyurt15Acta, zang, dag, yesilyurt16DetExp}.

Besides bipartite entangled states, multipartite entangled states can be used for quantum teleportation and superdense coding too.
What is more, being much more complicated than bipartite entanglement, multipartite entanglement introduces more diversity to the teleportation and superdense coding schemes.
Maximally entangled W states has been used in quantum teleportation protocols~\cite{gorbachev1,Joo}, but one needs to perform non-local operation to recover the unknown state~\cite{gorbachev1}.
That is to say, this limitation restrains the possibility to recover the unknown state using maximally entangled W states -so called \textit{prototype} W states-, resulting only imperfect teleportation schemes~\cite{Joo}.
Similar limitations arise on superdense coding with maximally entangled W states~\cite{agrawal}.
Because of these imperfections with W states, an intense effort have been devoted to finding a special class of W states which can enable perfect teleportation and superdense coding.
Gorbachev \emph{et al.} proposed a detailed scheme teleporting entangled states via the W-class state quantum channel~\cite{gorbachev}.
Agrawal \emph{et al.} showed that there exists a special class of three-qubit W states that can be used for perfect teleportation and superdense coding~\cite{agrawal}, and then Li \emph{et al.} generalized these schemes to the cases with W-class states in higher-dimension systems~\cite{li}.
This special class of W-like states (denoted by $|\mathcal{W}\rangle_N$ in this letter) can be used for perfect teleportation and superdense coding, so it is of great importance to generate $|\mathcal{W}\rangle_N$ states.

In this letter, we propose a scheme to prepare $|\mathcal{W}\rangle_N$ states via polarization dependent beam splitter (PDBS) based fusion and expansion mechanism for polarization encoded photons.
As a byproduct, a large-scale maximally entangled W state can be generated by fusing or expanding the $|\mathcal{W}\rangle_N$ states too, and the results show that this scheme is more efficient than the one starting from maximally entangled W states~\cite{like}.
This letter, is organized as follows: In Section II, we introduce the PDBS and the fusion or expansion strategies for $|\mathcal{W}\rangle_N$ states.
The strategies for expending or fusing $|\mathcal{W}\rangle_N$ states into large-scale maximally entangled states are presented in Section III.
In Section IV, we discuss the resource cost of the fusion strategy for $|\mathcal{W}\rangle_N$ states and the results are summarized in Section V.

\section{PDBS-based fusion and expansion strategies for $|\mathcal{W}\rangle_N$ states}
In this section, we show how to generate a large-scale $|\mathcal{W}\rangle_N$ state by fusing or expanding small-size $|\mathcal{W}\rangle_N$ states via PDBS.
PDBS has polarization-dependent transmissivities: $0<\mu<1$ for H polarization photons and $0<\nu<1$ for V polarization photons.
The function of the PDBS (as shown in red dashed rectangle in Fig.1) can be described by the following basic transformations~\cite{tashima1},
\begin{subequations}
\begin{eqnarray}
|H\rangle_a|H\rangle_b & \rightarrow & (2\mu-1)|H\rangle_ d|H\rangle_c + \sqrt{2\mu}\sqrt{1-\mu}|0\rangle_c|HH\rangle_d - \sqrt{2\mu}\sqrt{1-\mu}|HH\rangle_c|0\rangle_d, \\
|H\rangle_a|V\rangle_b & \rightarrow & \sqrt{\mu}\sqrt{\nu}|H\rangle_c|V\rangle_d - \sqrt{1-\mu}\sqrt{1-\nu}|V\rangle_c|H\rangle_d \nonumber\\
&+& \sqrt{1-\mu}\sqrt{\nu}|0\rangle_c|HV\rangle_d - \sqrt{\mu}\sqrt{1-\nu}|HV\rangle_c|0\rangle_d, \\
|V\rangle_a|H\rangle_b & \rightarrow & \sqrt{\mu}\sqrt{\nu}|V\rangle_c|H\rangle_d - \sqrt{1-\mu}\sqrt{1-\nu}|H\rangle_c|V\rangle_d \nonumber\\
&+& \sqrt{1-\mu}\sqrt{\nu}|0\rangle_c|VH\rangle_d - \sqrt{\mu}\sqrt{1-\nu}|VH\rangle_c|0\rangle_d, \\
|V\rangle_a|V\rangle_b & \rightarrow & (2\nu-1)|V\rangle_ d|V\rangle_c + \sqrt{2\nu}\sqrt{1-\nu}|0\rangle_c|VV\rangle_d - \sqrt{2\nu}\sqrt{1-\nu}|VV\rangle_c|0\rangle_d.
\end{eqnarray}
\end{subequations}

\begin{figure}[!htbp]
    \includegraphics[width=3.0in]{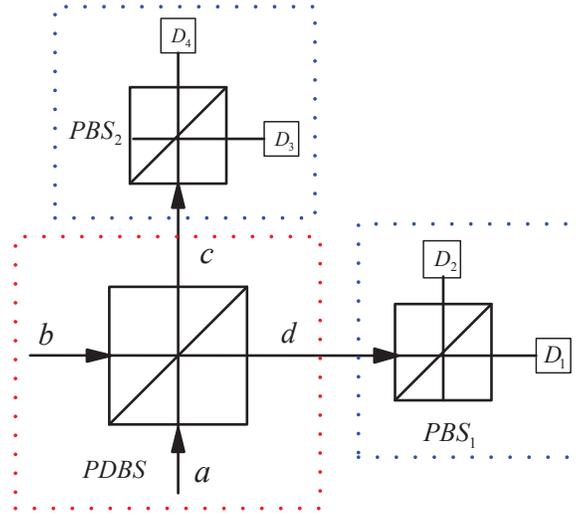}
    \label{fig1}
    \caption{The fusion and detection mechanism. The red-dashed rectangle presents the fusion mechanism, i.e. a PDBS. Two photons, one from $|\mathcal{W}\rangle_A$ and the other from $|\mathcal{W}\rangle_B$ are sent to the PDBS through input modes $a$ and $b$, respectively. The corresponding output modes are labeled as $c$ and $d$. The blue-dashed rectangles present the detection mechanisms. The output photon from the mode $c$($d$) is sent to the polarizing beam splitter (PBS), which reflects vertically (V) polarized photons and transmits horizontally (H) polarized photons, and the output modes of the PBSs are measured by detectors $D_1$, $D_2$, $D_3$ and $D_4$.}
\end{figure}

\subsection{Creation of large-scale $|\mathcal{W}\rangle$ state by fusing small-size $|\mathcal{W}\rangle$ states}
The special class of non-maximally entangled W states, which can be used for perfect teleportation and superdense coding is found as~\cite{li}:
\begin{eqnarray}
|\mathcal{W}\rangle_N=\frac{1}{\sqrt{2}}[|HH\ldots HV\rangle_{1\ldots N}+\frac{1}{\sqrt{N-1}}(|HH\ldots VH\rangle_{1\ldots N}+\ldots+|VH\ldots HH\rangle_{1\ldots N}],  (N\geq2).
\end{eqnarray}
To explain this class of W-like state, we first give the explicit expressions of prototype W states~\cite{ozdemir1}, i.e. $|W_n\rangle=\frac{1}{\sqrt{n}}[|({n-1})_H\rangle|1_V\rangle_1+\sqrt{n-1}|W_{n-1}\rangle|1_H\rangle_1]$.
A tripartite W state is written as $|W_3\rangle=\frac{1}{\sqrt{3}}(|HHV\rangle+|HVH\rangle+|VHH\rangle)=\frac{1}{\sqrt{3}}(|2_H\rangle|1_V\rangle_1+\sqrt{2}|W_2\rangle|1_H\rangle_1)$
with $W_2$ being the W-type Bell pair, i.e. $|W_2\rangle=\frac{1}{\sqrt{2}}(|HV\rangle+|VH\rangle)$ and obviously, bipartite $|\mathcal{W}\rangle$ state reduces to the $|W_2\rangle$ state.

Firstly, let's introduce the strategy for fusing a $|\mathcal{W}\rangle_2$ and a $|\mathcal{W}\rangle_3$ into a four-qubit  $|\mathcal{W}\rangle_4$ state.
Two initial W-class states can be written as
\begin{subequations}
\begin{eqnarray}
 |\mathcal{W}\rangle_2&=&\frac{1}{\sqrt{2}}[|H\rangle|V\rangle_a+|V\rangle|H\rangle_a],\\
 |\mathcal{W}\rangle_3&=&\frac{1}{\sqrt{2}}[|HH\rangle|V\rangle_b+\frac{1}{\sqrt{2}}(|HV\rangle|H\rangle_b+|VH\rangle|H\rangle_b)].
\end{eqnarray}
\end{subequations}
In the fusion strategy, we only have access to one photon of each W-class states, i.e. photons $a, b$. Photon $a$ is from $|\mathcal{W}\rangle_2$ and $b$ from  $|\mathcal{W}\rangle_3$, and they will be sent into the PDBS for completing the fusion process (as shown in Fig.1).
The remaining photons are kept intact at their sites. We are only interested in the case where a photon is present in each of the modes $c$ and $d$. So the state after the PDBS can be written as
\begin{eqnarray}
\frac{1}{2}[(2\nu-1)|HHH\rangle|V\rangle_c|V\rangle_d&-&\frac{1}{\sqrt{2}}\sqrt{1-\mu}\sqrt{1-\nu}|HHV\rangle|H\rangle_c|V\rangle_d+\frac{1}{\sqrt{2}}\sqrt{\mu}\sqrt{\nu}|HHV\rangle|V\rangle_c|H\rangle_d\nonumber\\
&-&\frac{1}{\sqrt{2}}\sqrt{1-\mu}\sqrt{1-\nu}|HVH\rangle|H\rangle_c|V\rangle_d+\frac{1}{\sqrt{2}}\sqrt{\mu}\sqrt{\nu}|HVH\rangle|V\rangle_c|H\rangle_d\nonumber\\
&-&\sqrt{1-\mu}\sqrt{1-\nu}|VHH\rangle|V\rangle_c|H\rangle_d+\sqrt{\mu}\sqrt{\nu}|VHH\rangle|H\rangle_c|V\rangle_d].
\end{eqnarray}
As shown in FIG. 1 (blue dashed rectangles), the detection mechanisms are placed after the output modes $c$ and $d$. It seems that the two photons are all detected during the fusion mechanism. Actually, only one photon detection is needed for the completion of the fusion process, and the other photon detection is only done for the verification of the event where a photon is present in each of the two output modes of the PDBS. As discussed in Ref.~\cite{tashima2}, the second photon detection can be replaced by a connection to a further optical circuit to complete further applications of the output $|\mathcal{W}\rangle_N$ state. That is to say, the verification process of the event where a photon is present in each of the two output modes of the PDBS and the applications of the output state are realized simultaneously. If a V-polarized photon is detected in $D_2$, a four-qubit W-like state will be generated as
\begin{eqnarray}
\frac{1}{2}C_1[|HHH\rangle|V\rangle_c+C_2|HHV\rangle|H\rangle_c+C_3|HVH\rangle|H\rangle_c+C_4|VHH\rangle|H\rangle_c],
\end{eqnarray}
with
\begin{eqnarray}
C_1&=&2\nu-1\nonumber,\\
C_2&=&C_3=-\frac{1}{\sqrt{2}}\frac{\sqrt{1-\mu}\sqrt{1-\nu}}{2\nu-1}\nonumber,\\
C_4&=&\frac{\sqrt{\mu}\sqrt{\nu}}{2\nu-1}.
\end{eqnarray}
Obviously, we can obtain a $|\mathcal{W}\rangle_4$ with $|C_2|=|C_3|=|C_4|=\frac{1}{\sqrt{3}}$, which means that the parameter $\nu$ for the PDBS must satisfy $4\nu^{3}+3\nu^{2}-6\nu+1=0$.
So, we can get the corresponding values of $(\nu, \mu)$, and the success probability $Ps(|\mathcal{W}\rangle_4)=(2\nu-1)^2/2$.
It should be noted that $(\nu, \mu)$ are defined in the range $(0, 1)$.
Thus the values of ($\nu$, $\mu$) are $(0.7726, 0.1283)$ or $(0.1890, 0.6823)$.
As $\nu$ has two values, so does the success probability $Ps(|\mathcal{W}\rangle_4)$.
Therefore, $Ps(|\mathcal{W}\rangle_4)$ can be maximized by choosing one set ($\nu$, $\mu$) from the two possible sets.

This scheme can be generalized to the case of fusing $|\mathcal{W}\rangle_N$ state and $|\mathcal{W}\rangle_M$ state into a $|\mathcal{W}\rangle_{N+M-1}$ state $(N\neq M)$.
The multi-qubit state $|\mathcal{W}\rangle_N$ can be used as a shared resource for teleportation and superdense coding, which was detailed in Ref.~\cite{li}.
Two initial states can be written as follows:
\begin{subequations}
\begin{eqnarray}
|\mathcal{W}\rangle_N=\frac{1}{\sqrt{2}}[|HH\ldots HV\rangle_{1\ldots N}+\frac{1}{\sqrt{N-1}}(|HH\ldots VH\rangle_{1\ldots N}+\ldots+|VH\ldots HH\rangle_{1\ldots N}], \\
|\mathcal{W}\rangle_M=\frac{1}{\sqrt{2}}[|HH\ldots HV\rangle_{1\ldots M}+\frac{1}{\sqrt{M-1}}(|HH\ldots VH\rangle_{1\ldots M}+\ldots+|VH\ldots HH\rangle_{1\ldots M}].
\end{eqnarray}
\end{subequations}
Two photons coming from these two initial states (say $N$th and $M$th photons) respectively, will be sent into the input modes $a$ and $b$ of the PDBS.
Similarly, we are only interested in the case where a photon is present in each of the modes $c$ and $d$, and the conditions the parameters $\mu, \nu$ must satisfy can be written as
\begin{subequations}
\begin{eqnarray}
4(N-M)\nu^{3}+(-9N+3M+6)\nu^{2}+(6N-6)\nu-(N-1)=0,
\end{eqnarray}
\begin{eqnarray}
\frac{\sqrt{\mu\nu}}{\sqrt{N-1}}=\frac{\sqrt{(1-\mu)(1-\nu)}}{\sqrt{M-1}}.
\end{eqnarray}
\end{subequations}
By fixing $N$ and $M$, one can get the corresponding values of $\mu$ and $\nu$ by solving the two equations above.
However, with the increase of the size of the output state $|\mathcal{W}\rangle_k (k\geq 5)$, there are more and more solutions.
For example, $|\mathcal{W}\rangle_5$ can be generated by fusing two $|\mathcal{W}\rangle_3$ states or a $|\mathcal{W}\rangle_2$ and a $|\mathcal{W}\rangle_4$ state.
But according to the discussions in Ref.~\cite{ozdemir1}, there is such a conclusion that the closer the sizes of the two resource states are, the lower the cost is, i.e. it is optimal to
generate $|\mathcal{W}\rangle_N$ states by fusing two resource states of similar size.
Following this approach, in Table.\ref{table1}, we give a list of the values for $N, M, \mu, \nu$ and $Ps(|\mathcal{W}\rangle_{N+M-1})$ for some optimal fusion strategies.
$Ps(|\mathcal{W}\rangle_{N+M-1})$ is the corresponding maximum success probability ($(2\nu-1)^2/2$) for the combinations $(\nu, \mu)$. Moreover, it should be emphasized that the values of ($\nu$, $\mu$) do not depend on the size of the input states when the size of the input states are equal, i.e. when fusing two identical $|\mathcal{W}\rangle_N$ states, the values of ($\nu$, $\mu$) are fixed $\mu=(3\pm\sqrt{3})/6, \nu=(3\mp\sqrt{3})/6$.

By using this scheme, $|\mathcal{W}\rangle_3$ state can be prepared by fusing two $|\mathcal{W}\rangle_2$ states, and notice that the $|\mathcal{W}\rangle_2$ state is a maximally entangled Bell state. So if we take the Bell state as the initial resource, arbitrary size $|\mathcal{W}\rangle_N$ can be generated via the fusion strategy above.

In addition, if an H-polarized photon is detected in $D_1$, the fusion process fails in general ($N >2,M > 2$). In particular, when $N=M=2$ , $|\mathcal{W}\rangle_3$ state will be generated if V-polarized photon( H-polarized photon) is detected in $D_2$ ($D_1$). It is worth mentioning that the Bell state can also be generated from two single-photon states by this fusion strategy too~\cite{like}.

\begin{table}[t]
    \caption{List of the values of $N, M, \mu, \nu$ and $Ps(|\mathcal{W}\rangle_{N+M-1})$.}
    \begin{tabular}{c|c|c|c|cc} \hline\hline
 N        &       M       &       $(\nu, \mu)_1$       &      $(\nu, \mu)_2$       &     $Ps(\widetilde{W}^{N+M-1})_{max}$    & \\
 \hline
 2        &       2       &     $(0.7887, 0.2113)$     &    $(0.2113, 0.7887)$     &                    0.1667                & \\
 \hline
 2        &       3       &     $(0.7726, 0.1283)$     &    $(0.4890, 0.6823)$     &                    0.1486                & \\
 \hline
 3        &       3       &     $(0.7887, 0.2113)$     &    $(0.2113, 0.7887)$     &                    0.1667                & \\
 \hline
 3        &       4       &     $(0.7789, 0.1598)$     &    $(0.1990, 0.7285)$     &                    0.1812                & \\
 \hline
 4        &       4       &     $(0.7887, 0.2113)$     &    $(0.2113, 0.7887)$     &                    0.1667                & \\
 \hline
 4        &       5       &     $(0.7812, 0.1735)$     &    $(0.2028, 0.7467)$     &                    0.1767                & \\
 \hline
 5        &       5       &     $(0.7887, 0.2113)$     &    $(0.2113, 0.7887)$     &                    0.1667                & \\
 \hline
 5        &       6       &     $(0.7828, 0.1816)$     &    $(0.2047, 0.7573)$     &                    0.1744                & \\
 \hline
 6        &       6       &     $(0.7887, 0.2113)$     &    $(0.2113, 0.7887)$     &                    0.1667                & \\
 \hline
 6        &       7       &     $(0.7838, 0.1868)$     &    $(0.2060, 0.7629)$     &                    0.1729                & \\
 \hline
 7        &       7       &     $(0.7887, 0.2113)$     &    $(0.2113, 0.7887)$     &                    0.1667                & \\
\hline
 7        &       8       &     $(0.7846, 0.1906)$     &    $(0.2069, 0.7665)$     &                    0.1718                & \\
\hline
 8        &       8       &     $(0.7887, 0.2113)$     &    $(0.2113, 0.7887)$     &                    0.1667                & \\
\hline
 8        &       9       &     $(0.7851, 0.1932)$     &    $(0.2075, 0.7697)$     &                    0.1711                & \\
\hline
 9        &       9       &     $(0.7887, 0.2113)$     &    $(0.2113, 0.7887)$     &                    0.1667                & \\
\hline
 9        &       10      &     $(0.7855, 0.1953)$     &    $(0.2080, 0.7716)$     &                    0.1705                & \\
\hline
 10       &       10      &     $(0.7887, 0.2113)$     &    $(0.2113, 0.7887)$     &                    0.1667                & \\
\hline
 .        &       .       &             .              &            .              &                       .                  & \\
 .        &       .       &             .              &            .              &                       .                  & \\
 .        &       .       &             .              &            .              &                       .                  & \\
\hline\hline
\end{tabular}
     \label{table1}
\end{table}

\subsection{Creation of large-scale $|\mathcal{W}\rangle$ state by expanding small-size $|\mathcal{W}\rangle$ states}

Here, we propose a simple scheme for expanding a polarization-entangled $|\mathcal{W}\rangle_N$ state by adding an H-polarized ancilla photon.
With the help of a PDBS,  one of the photons in the $|\mathcal{W}\rangle_N$ state will interfere with an H-polarized photon, and after post-selection a $|\mathcal{W}\rangle_{N+1}$ state can be generated.
As depicted in Fig.1
one (say the $N$th) of the photons in the $|\mathcal{W}\rangle_N$ state is inputted in mode $a$, and the H-polarized auxiliary photon is added in mode $b$.
Here, we only select those events where there is only one photon in each output mode.
After coupling on the PDBS, the state of the photons can be written as
\begin{eqnarray}
\frac{1}{\sqrt{2}}[C_1|HH\ldots HVH\rangle_{1\ldots N (N+1)}+C_2|HH\ldots HHV\rangle_{1\ldots N (N+1)}\nonumber \\
+C_3(|HH\ldots VHH\rangle_{1\ldots N (N+1)}+\ldots+|VH\ldots HHH\rangle_{1\ldots N (N+1)}],
\end{eqnarray}
with
\begin{eqnarray}
C_1&=&\sqrt{\mu}\sqrt{\nu}\nonumber,\\
C_2&=&=-\sqrt{(1-\mu)(1-\nu)}\nonumber,\\
C_3&=&\frac{(2\mu-1)}{\sqrt{N-1}}.
\end{eqnarray}
In addition, when $|\frac{C_1}{C_2}|=|\frac{C_3}{C_2}|=\frac{1}{\sqrt{N}}$, a $|\mathcal{W}\rangle_{N+1}$ state will be obtained with success probability  $Ps(|\mathcal{W}\rangle_{N+1}) = (1-\mu)(1-\nu)$, and the corresponding condition that $\mu$ must satisfy is $4(N-1)\mu^3-(3N-7)\mu^2-4\mu+1=0$. The values of $\mu$ and $\nu$ will be determined when N is given. For example, as $N=2$, $\mu$ and $\nu$  have two possible combinations: $\mu_1=0.2991$, $\nu_1=0.5398$ and $\mu_2=0.6799$, $\nu_2=0.1904$. In this case, no photon detection is needed during the expansion process, and the verification process of the event where a photon is present in each of the two output modes of the PDBS and the applications of the output state are realized simultaneously.

\section{Creation of large-size maximally entangled W states by fusing or expanding $|\mathcal{W}\rangle$ states}
In this section, we propose an effective scheme to prepare large-scale maximally entangled W states by fusing or expanding $|\mathcal{W}\rangle_N$ states.
Based on the selective transmission rates of PDBS for different polarization states, a maximally entangled state can be generated by selecting suitable parameters $(\mu,\nu)$.
Because the scheme can generate maximally entangled states in terms of non-maximally entangled states, it may play important roles in quantum communication.

\subsection{Creation of large-scale $|W\rangle$ state by fusing small-size $|\mathcal{W}\rangle$ states}
We will now demonstrate how a large-scale maximally entangled W state can be generated by fusing small-size $|\mathcal{W}\rangle_N$ states.
First, let's consider the case of fusing two identical tri-photon W-class states into a five-photon maximally entangled state.
Two input states can be written as
\begin{subequations}
\begin{eqnarray}
|\mathcal{W}\rangle_3=\frac{1}{\sqrt{2}}[|HHV\rangle_{123}+\frac{1}{\sqrt{2}}(|VHH\rangle_{123}+|HVH\rangle_{123})],\\
|\mathcal{W}\rangle_3=\frac{1}{\sqrt{2}}[|HHV\rangle_{456}+\frac{1}{\sqrt{2}}(|VHH\rangle_{456}+|HVH\rangle_{456})].
\end{eqnarray}
\end{subequations}
In this fusion process, two photons (say photons 3, 6) respectively coming from these two $|\mathcal{W}\rangle_3$ states will be sent into the input modes $a$ and $b$ of a PDBS. In the case where there is only one photon in each output mode, we can get a five-qubit W-like state:
\begin{eqnarray}
|\widetilde{W}\rangle_5=C_1|HHHH\rangle_{1245}|V\rangle_c+C_2|HHHV\rangle_{1245}|H\rangle_c+C_3|HHVH\rangle_{1245}|H\rangle_c\nonumber\\
+C_4|HVHH\rangle_{1245}|H\rangle_c+C_5|VHHH\rangle_{1245}|H\rangle_c,
\end{eqnarray}
with
\begin{eqnarray}
C_1&=&\frac{1}{2}(2\nu-1)\nonumber, \\
C_2&=&-\frac{1}{2\sqrt{2}}\sqrt{1-\mu}\sqrt{1-\nu}\nonumber, \\
C_3&=&-\frac{1}{2\sqrt{2}}\sqrt{1-\mu}\sqrt{1-\nu}\nonumber, \\
C_4&=&\frac{1}{2\sqrt{2}}\sqrt{\mu}\sqrt{\nu}\nonumber, \\
C_5&=&\frac{1}{2\sqrt{2}}\sqrt{\mu}\sqrt{\nu},
\end{eqnarray}
 when a V-polarized photon is detected in detector $D_2$. If the parameters $(\mu, \nu)$ for the PDBS are chosen to be $\mu=2/3, \nu=1/3$ or vice versa,  $C_1=C_2=C_3=C_4=C_5$, and thus the W-like state obtained here becomes a standard W state. Then a five-photon maximally entangled state can be generated with success probability $Ps(W_5)=5/36$.

This fusion scheme can be generalized to the case of generating a $(2N-1)$-qubit maximally entangled W state by fusing two $|\mathcal{W}\rangle_N$ states.
To begin with this fusion strategy, one (say the $N$th) photon from each $|\mathcal{W}\rangle_N$ state will be sent into a PDBS and the parameters $(\mu, \nu)$ for the PDBS are chosen to be $\mu=[(4N-3)+\sqrt{4N-3}]/2(4N-3), \nu=[(4N-3)-\sqrt{4N-3}]/2(4N-3)$ or vice versa. If we are only interested in the case where a photon is present in each of the output modes, a maximally entangled $|W\rangle_{2N-1}$ state can be generated when a V-polarized photon is detected in $D_2$:
\begin{eqnarray}
|W\rangle_{2N-1}&=&\frac{1}{\sqrt{2N-1}}[|HH\ldots HV\rangle_{12\ldots (2N-1)}+|HH\ldots VH\rangle_{12\ldots (2N-1)}+\ldots \nonumber\\
&+&|HV\ldots HH\rangle_{12\ldots (2N-1)}+|VH\ldots HH\rangle_{12\ldots (2N-1)}].
\end{eqnarray}
The fusion process is also applicable to fusing a $|\mathcal{W}\rangle_N$ state and a $|\mathcal{W}\rangle_M$ state ( $N\neq M$) into a maximally entangled $|W\rangle_{N+M-1}$ state.
In such a situation, the values of $\mu, \nu$ are dependent on $N, M$:
\begin{subequations}
\begin{eqnarray}
4(N-M)\nu^{3}&+&(4M-8N+3)\nu^{2}+(5N-M-3)\nu-(N-1)=0,\\
\nonumber\\
\frac{\sqrt{\mu\nu}}{\sqrt{N-1}}&=&\frac{\sqrt{(1-\mu)(1-\nu)}}{\sqrt{M-1}}.
\end{eqnarray}
\end{subequations}
When $N, M$ are given, $\mu$ and $\nu$ can be determined.
For example, when $N=3$ and $ M=4$, $(\nu, \mu)$  have two possible combinations: $(0.3448, 0.5589)$ or $(0.6469, 0.2669)$.
Here $(\nu, \mu)$ are defined in the range $(0, 1)$ too, and the success probability is $Ps(W_{N+M-1})=(2\nu-1)^2(N+M-1)/4$.

It should be emphasized that, although we only discussed the situation of the photon is present in each of the modes $c$ and $d$, we also explored the case where two photons are in the same output mode. As in Ref. \cite{li}, there exists recyclable case in our strategy, i.e., two maximally entangled $W$ states $|W_{n-1}\rangle$ and $|W_{m-1}\rangle$ can be left if two $H$-polarized photons are detected in $D_1$. According to Ref. \cite{li}, these two maximally entangled $W$ states can serve as the initial resource for the further fusion.

\subsection{Creation of large-scale $|W\rangle$ state by expanding small-size $|\mathcal{W}\rangle$ states}
In this subsection, let's introduce the strategy for expanding $|\mathcal{W}\rangle_N$ state into maximally entangled $|W_{N+1}\rangle$ state.
The $|\mathcal{W}\rangle_N$ state can be written as in Eq.$(2)$.
One of the photons (say the $N$th) from $|\mathcal{W}\rangle_N$ and an H-polarized auxiliary photon will be sent into a PDBS for completing the expansion process.
If we are only interested in the case where a photon is present in each of the output modes of the PDBS, the following state can be generated:
\begin{eqnarray}
\frac{1}{\sqrt{2}}[C_1|HH\ldots HVH\rangle_{1\ldots N (N+1)}+C_2|HH\ldots HHV\rangle_{1\ldots N (N+1)}\nonumber\\
+C_3(|HH\ldots VHH\rangle_{1\ldots N (N+1)}+\ldots+|VH\ldots HHH\rangle_{1\ldots N (N+1)})].
\end{eqnarray}
with
\begin{eqnarray}
C_1&=&\sqrt{\mu}\sqrt{\nu}\nonumber,\\
C_2&=&=-\sqrt{(1-\mu)(1-\nu)}\nonumber,\\
C_3&=&\frac{(2\mu-1)}{\sqrt{N-1}}.
\end{eqnarray}

If the parameters $(\mu, \nu)$ for the PDBS are chosen to be $\mu=[(N+3)\pm\sqrt{(N+3)(N-1)}]/2(N+3), \nu=[(N+3)\mp\sqrt{(N+3)(N-1)}]/2(N+3)$, $|C_1|=|C_2|=|C_3|$ holds, which means that when the input states are known for us, a maximally entangled state can be generated with success probability $Ps(W_{N+1})=\mu\nu(N+1)/2$ by ajusting the parameters $(\mu, \nu)$ for the PDBS.

\section{Cost analysis and discussion}
In this section, we will make a brief discussion on the resource cost of our strategies for fusing small-size $|\mathcal{W}\rangle_N$ states into larger-scale $|\mathcal{W}\rangle_N$ state or W state.
With the PDBS fusion mechanism, a large-scale maximally entangled W state can be generated by fusing small-size $|\mathcal{W}\rangle_N$ states.
In addition, one of our previous works shows that the PDBS fusion mechanism can be used to generate a large-scale maximally entangled W state by fusing small-size maximally entangled W states too~\cite{like}.
So it is necessary to make a cost comparison analysis among these two strategies.
Because the two fusion strategies both start from $|W_2\rangle$ state, we can define the basic resource cost for preparing $|W_2\rangle$  as the unit cost, i.e. $R[W_2]=1$, and make a comparison on the optimal cost of these two strategies by numerical evaluation.

\subsection{Cost analysis of fusing small-size $|\mathcal{W}\rangle_N$ states into larger-scale $|\mathcal{W}\rangle_N$ state}
In Ref.~\cite{ozdemir1}, the resource cost of preparing a large W state is defined as follows:
\begin{eqnarray}
R[W_{m+n-1}] = \frac{R[W_m]+R[W_n]}{P_s(W_m,W_n)},
\end{eqnarray}
where both $W_m$ and $W_n$ are maximally entangled W states, and $P_s(W_m,W_n)$ is the success probability of the fusion process.
Similarly, because our fusion strategy starts from the standard $|W_2\rangle$ state, the resource cost of generating a $|\mathcal{W}\rangle_{N+M-1}$ state by fusing $|\mathcal{W}\rangle_N$ state and a $|\mathcal{W}\rangle_M$ state can also be calculated in the same way.
The success probability of getting a $|\mathcal{W}\rangle_{N+M-1}$ state is $Ps(|\mathcal{W}\rangle_{N+M-1})= (2\nu-1)^2/2$, and thus the resource cost can be expressed as
\begin{eqnarray}
R[\mathcal{W}_{N+M-1}]=\frac{(R[\mathcal{W}_N]+R[\mathcal{W}_M])}{Ps(\mathcal{W}_{N+M-1})}.
\end{eqnarray}
The numerical results of resource cost $R[\mathcal{W}_{N+M-1}]$ are shown in Fig.2
for the optimal fusion processes, where the sizes of the input state are similar and the parameters of the PDBS have been chosen to maximize the success probabilities.
\begin{figure}[!htbp]
    \includegraphics[width=4.0in]{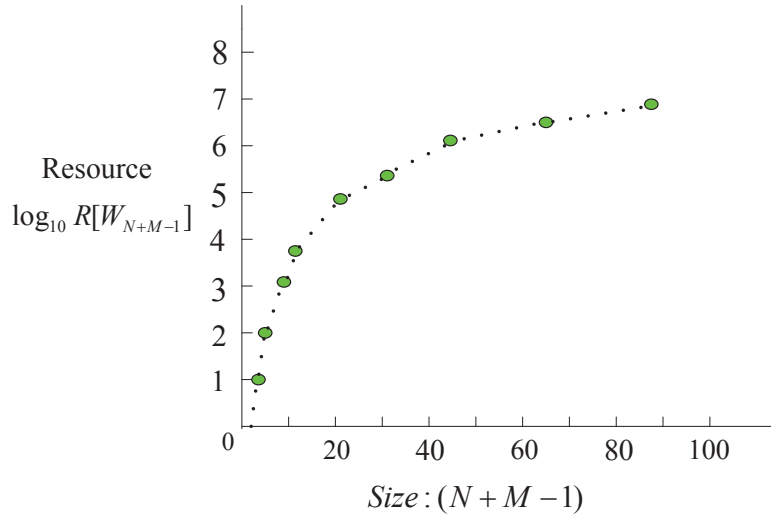}
    \label{fig2}
    \caption{(Color online) The resource costs of fusing a $|\mathcal{W}\rangle_N$ state and a $|\mathcal{W}\rangle_M$
     state into a $|\mathcal{W}\rangle_{N+M-1}$ state. The basic resource is the maximally entangled $|W\rangle_2$ state.
     The vertical axis is the resource cost and horizontal axis is the size of the output $|\mathcal{W}\rangle$ state.}                                                                                                                                             \end{figure}

\subsection{Cost comparison}
Here, we compare the resource cost of two different strategies for generating a maximally entangled W state.
In the first strategy, a large-scale maximally entangled W state can be generated by fusing two small-size maximally entangled W states (a $|W_n\rangle$ state and a $|W_m\rangle$ state)~\cite{like}, and its resource cost is denoted as $R[W_{m+n-1}]$.
The second fusion is the one we proposed in this letter, i.e. a large-scale maximally entangled $|W_{n+m-1}\rangle$ state can be generated by fusing a $|\mathcal{W}_n\rangle$ state and a $|\mathcal{W}_m\rangle$ state, and its resource cost is denoted by $R[W_{m+n-1}]^{\prime}$.
Both $R[W_{m+n-1}]$ and $R[W_{m+n-1}]^{\prime}$ are calculated in terms of the unit $R[W_2]=1$.
FIG.3 shows that the resource cost $R[W_{m+n-1}]^{\prime}$ of the second method (red) is lower than that of the one using the first method (black). The fusion strategy starting from nonmaximally entangled $|\mathcal{W}\rangle$ states is more efficient than the one starting from maximally entangled W states.
\begin{figure}[!htbp]

    \includegraphics[width=4.0in]{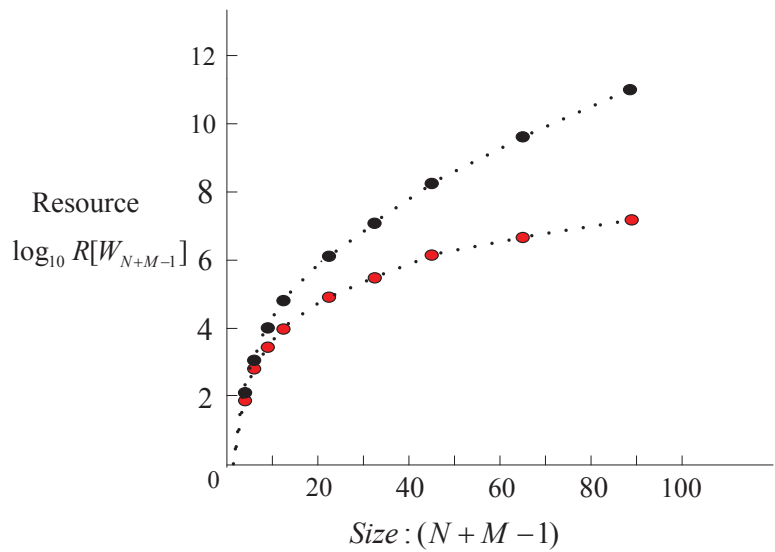}
    \label{fig3}
    \caption{(Color online) The resource costs comparison: the fusion strategy of generating $|W_{N+M-1}\rangle$ from $|W_n\rangle$ and $|W_m\rangle$ (black), and the fusion strategy of generating $|W_{N+M-1}\rangle$  from $|\mathcal{W}\rangle_N$ and $|\mathcal{W}\rangle_M$ (red). The basic resource cost for preparing maximally entangled $|W_2\rangle$  is defined as unit cost. The vertical axis is the resource cost and horizontal axis is the size of the output W state.}
                                                                                                             \end{figure}

\section{Conclusion}
In conclusion, we have proposed a fusion strategy of generating large-scale $|\mathcal{W}\rangle$ photonic state, which plays an important role in perfect teleportation and superdense coding.
Our fusion strategy can start from Bell states.
The main fusion mechanism is a PDBS, and no controlled gate is needed.
The fusion process can succeed by adjusting the parameters of the PDBS, which makes the current scheme simple and feasible.
Furthermore, with this fusion mechanism, a large-scale maximally entangled W state can be generated by fusing small-size $|\mathcal{W}\rangle$ states as well.
The cost analysis show that our fusion strategy is more efficient than the one starting from maximally entangled W states~\cite{like}.
The possibilities of generating large-scale $|\mathcal{W}\rangle$ states and $|W\rangle$ states via expansion mechanism have also been studied.

\begin{acknowledgments}
This work is supported by the National Natural Science Foundation of China (NSFC) under Grant Nos.11274010, 11204002, 11374085, 11204061 and  61370090; the Specialized Research Fund for the Doctoral Program of Higher Education (Grants No.20113401110002 and No.20123401120003); the personnel department of Anhui Province. F. Ozaydin and M. Yang are funded by Isik University Scientific Research Funding Agency under Grant Number: BAP-14A101.
\end{acknowledgments}

\end{document}